\documentclass{sf2a-conf2013}
\usepackage{graphicx}
\usepackage{hyperref}
\usepackage[]{natbib}  
\usepackage{epstopdf}

\def\BibTeX{{\rm B\kern-.05em{\sc i\kern-.025em b}\kern-.08em
    T\kern-.1667em\lower.7ex\hbox{E}\kern-.125emX}}
\bibpunct{(}{)}{;}{a}{}{,}  

\begin{document}

\TitreGlobal{SF2A 2013}

\title{Dust in the wind II:\\ Polarization imaging from disk-born outflows}

\runningtitle{Dust in the wind II}

\author{F. Marin$^*$}\address{Observatoire Astronomique de Strasbourg, Universit\'e de Strasbourg, 
			      CNRS, UMR 7550, 11 rue de l'Universit\'e, 67000 Strasbourg, France\\}
		 \thanks{$^*$ frederic.marin@astro.unistra.fr}

\author{R. W. Goosmann$^1$}

\setcounter{page}{237}

\index{Marin, F.}
\index{Goosmann, R. W.}


\maketitle

\begin{abstract}
In this second research note of a series of two, we aim to map the polarized flux emerging from a disk-born, dusty outflow as it was prescribed by \citet{Elvis2000}.
His structure for quasars was achieved to unify the emission and absorption features observed in active galactic nuclei (AGN) and can be used as an alternative
scenario to the typical dusty torus that is extensively used to account for AGN circumnuclear obscuration. Using Monte Carlo radiative transfer simulations, we
model an obscuring outflow arising from an emitting accretion disk and examine the resulting polarization degree, polarization angle and polarized flux.
Polarization cartography reveals that a disk-born outflow has a similar torus morphology in polar viewing angles, with bright polarized fluxes reprocessed
onto the wind funnel. At intermediate and edge-on inclinations, the model is rather close to a double-conical wind, with higher fluxes in the cone bases. 
It indicates that the optically thick outflow is not efficient enough to avoid radiation escaping from the central region, particularly due to the geometrically 
thin divergence angle of the outflow. As parametrized in this research note, a dusty outflow does not seem to be able to correctly reproduce the polarimetric
behavior of an usual dusty torus. Further refinement of the model is necessary.
\end{abstract}



\section{Introduction}
The anisotropic obscuration of the central region of AGN and quasars is well explained by the unified scheme \citep{Antonucci1993,Urry1995}. In this scenario,
the fueling engine of AGN (a supermassive black hole SMBH and its surrounding, reprocessing accretion disk) is hidden along the equatorial plane by a circumnuclear medium 
situated at a few parsecs from the SMBH \citep{Krolik1986,Krolik1988}. Illuminated by the continuum source, the obscuring material is expected to re-emit in 
the infrared domain, as the UV-heated dust particles re-emit at lower wavelengths. Such hypotheses were strongly supported by early photometric infrared 
observations of Seyfert-like galaxies \citep{Low1968,Kleinmann1970,Kleinmann1970b}, and an indubitable confirmation was found by \citet{Jaffe2004} and 
\citet{Wittkowski2004}. They observed the dusty heart of NGC~1068 using mid-infrared interferometry and resolved the outer morphology of the obscuring
region. The resulting morphology is close to a toroidal configuration; a geometry usually adopted in many AGN simulations 
\citep{Kartje1995,Wolf1999,Young2000,Watanabe2003,Goosmann2007,Marin2012}. However, the actual morphology of the circumnuclear gas surrounding the AGN 
equator is highly debated. While simple, torus-like models give a sufficient approximation for radiative transfer simulations, more complex structures are 
actually investigated (see \citealt{Elitzur2006} for a review). 

This research note focuses on the peculiar model given by \citet{Elvis2000,Elvis2012}, where the dusty torus is pictured as an 
optically thick, equatorial outflow originating close to the supermassive black hole. While the presence of dust grains at a hundred of gravitational
radii seems uncommon, \citet{Czerny2012} proved that dust can be formed in accretion disk atmospheres. The dust mixture is protected by an internal,
shielding outflow composed of a flow of warm, highly ionized matter (WHIM) and can rise from the disk atmosphere, before being evaporated at high altitudes.
It results in a dust outflow that propagates mainly along the equatorial plane, an interesting configuration to explain the edge-on obscuration of AGN 
\citep{Antonucci1985,Pier1992}.

Using an academic model inspired by our previous analyses (Marin \& Goosmann, accepted) of the model proposed by \citet{Elvis2000}, we aim to produce polarization
maps of a dusty outflow in order to compare it with observations. This research note is the second in a series of two, both included in this volume. Polarization spectra
were investigated in the first issue (Marin \& Goosmann, submitted).

\section{Model setup}
The empirically-derived structure for quasars presented by \citet{Elvis2000} stipulates that a flow of matter arises from a narrow range of radii on the
accretion disk. The material is bent and accelerated outward by internal radiation pressure, with an angle $\theta$ = 60$^\circ$ and a divergence angle
$\delta\theta$ = 3$^\circ$. This angular parametrization is given by the ratio of narrow absorption line (NAL) to absorption-free AGN 
\citep{Reynolds1997,Crenshaw1999}, while the morphological parameters (distance to the ionizing source r$_1$, width of the column flow r$_2$) are derived 
from constraints of the broad emission line (BEL) region \citep{Peterson1997}. To follow the prescription of \citet{Elvis2000} and our previous investigation 
(Marin \& Goosmann, submitted), we set r$_1$ to 0.0032~pc (10$^{16}$~cm) and r$_2$ to 0.00032~pc (10$^{15}$~cm). The wind extends up to r$_3$~=~0.032~pc
(10$^{17}$~cm). The dust mixture, filling our model, is taken from \citet{Wolf1999}. We opted for an optically thick outflow, with optical depths
$\tau_2 \sim$~3600 along the wind and $\tau_1 \sim$~36 along the equator. The model is summarized in Marin \& Goosmann (submitted, Fig.~1).

We use the radiative transfer Monte Carlo code {\sc stokes} \citep{Goosmann2007}, upgraded with a polarization imaging technic \citep{Marin2012,Marin2012b} 
to compute the net polarization emerging from the model. The unpolarized, input spectrum comes from an isotropic, disk-like emitting region and has a power-law 
spectral energy distribution $F_{\rm *}~\propto~\nu^{-\alpha}$ and $\alpha = 1$.

\section{Wavelength-integrated polarization maps}
\begin{figure}
   \centering
   \begin{tabular}{ll}
      \includegraphics[trim = 0mm 0mm 0mm 10mm,width=0.48\textwidth,clip]{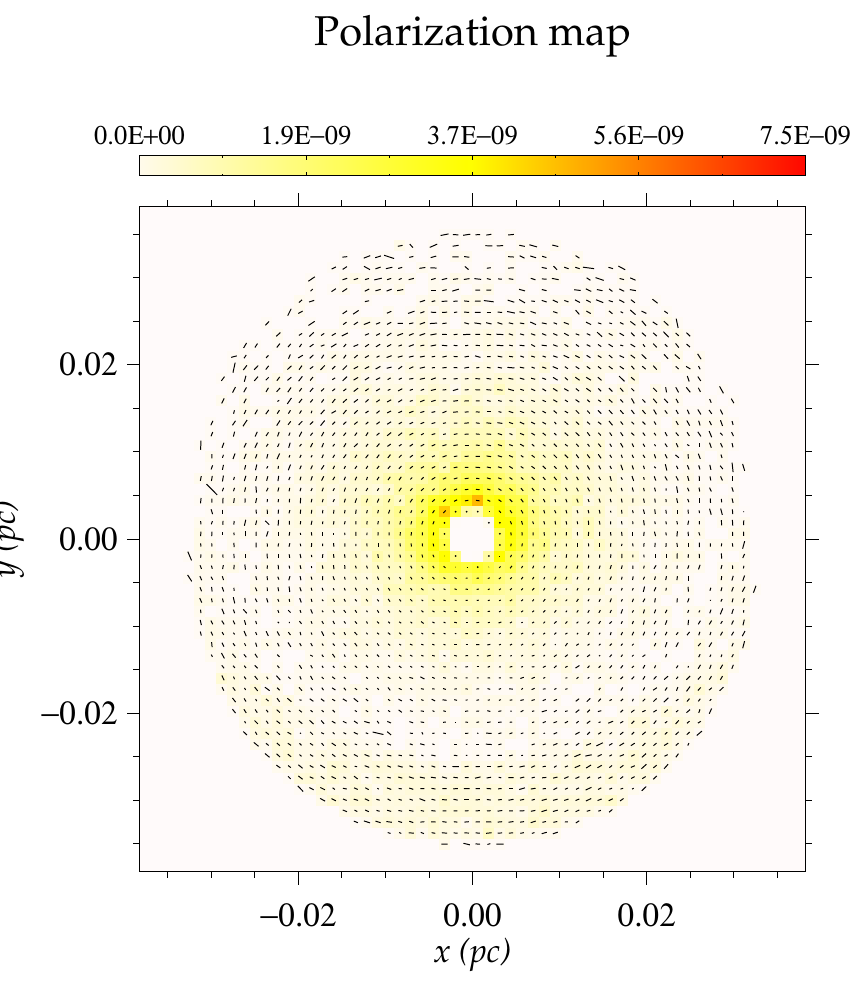} 
      \hspace{0.02\textwidth}
      \includegraphics[trim = 0mm 0mm 0mm 10mm,width=0.48\textwidth,clip]{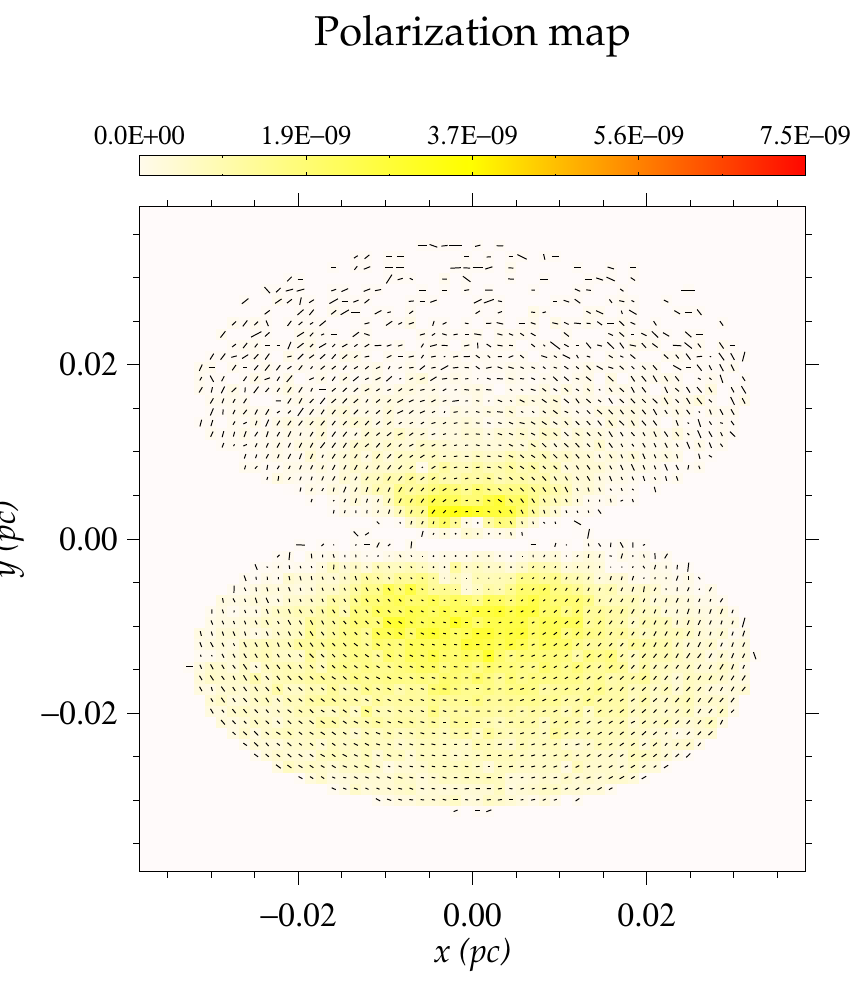} \\
   \end{tabular}
      \includegraphics[trim = 0mm 0mm 0mm 10mm,width=0.48\textwidth,clip]{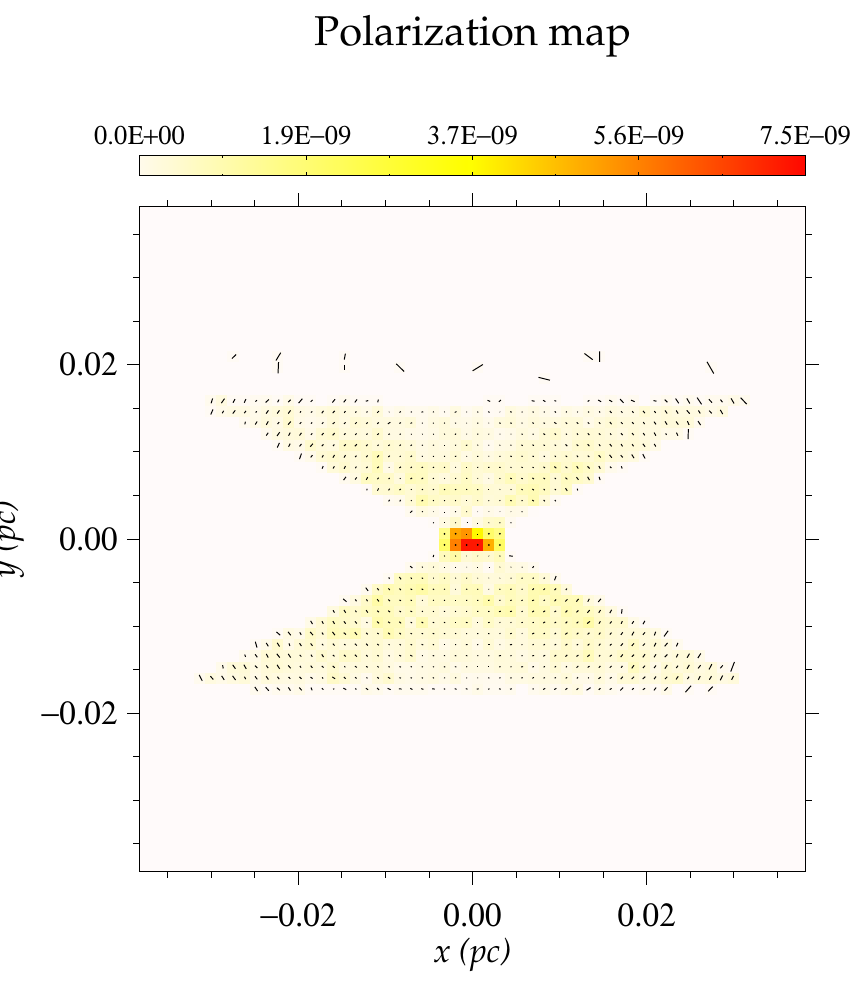} 
      \caption{Modeling the polarized flux, $PF_{\nu}$, induced by complex reprocessing in
	       the structure for quasar \citep{Elvis2000}. $PF_{\nu}$ is color-coded and 
	       integrated over the 2000 -- 8000~\AA~band.
	       \textit{Top-left}: face-on image;
	       \textit{Top-right}: image at $i \sim 60^\circ$;
	       \textit{Bottom}: edge-on image.}
     \label{Fig1}%
\end{figure}

In Fig.~\ref{Fig1}, we present the simulated polarization cartography of the dusty model by \citet{Elvis2000}. The maps simultaneously show the polarized flux, 
$PF_{\nu}$, the polarization percentage $P$, and the polarization position angle, $\psi$. The angle $\psi$ is represented by black bars drawn in the center 
of each spatial bin and the length of the vector is proportional to $P$. A vertical bar indicates a polarization of $\psi$ = 90$^\circ$, a bar leaning to 
the right denotes 90$^\circ > \psi > 0^\circ$ and a horizontal bar stands for $\psi$ = 0$^\circ$. For each pixel, the Stokes 
parameters are integrated over the full 2000 -- 8000~\AA~ range.

At a polar viewing angle (Fig.~\ref{Fig1}, top-left), the funnel of the outflow is illuminated by the central source, and multiple forward and backward scattering 
events are responsible for the high polarized flux detected in the vicinity of the emitting region. The illumination decreases with distance from the source and 
nearly no polarized flux is detected at the most extended parts of the outflows. The spectropolarimetric pattern is similar to the polarization maps produced for an isolated 
dusty torus \citep{Marin2012} and appears not to be distinguishable by polarization imaging.

In the intermediate inclination case (Fig.~\ref{Fig1}, top-right), the flux mainly comes from backscattering events on the lower part of the model, where photons 
escaping from the geometrically thin outflow ($\delta\theta$ = 3$^\circ$) are reprocessed. Due to the important optical depth along the wind flow, all of the 
polarized flux is absorbed at the wind base. The overall shape of the model is that of a double-cone, with a maximum polarized flux detected in the wind bases.
However, the flux gradient, as seen from previous modeling of hourglass-shaped outflows \citep{Marin2012}, is absent due to the hollow geometry of the model.

Finally, along the equatorial viewing angle (Fig.~\ref{Fig1}, bottom), the outflow is nearly invisible as it absorbs most of the incident photon flux. However, a 
large fraction of radiation can escape by transmission through the equatorial medium and collaborate to rise the polarization degree detected in type-2 viewing 
angles. A few photons are detected on the northern part of the map, due to rare backscattering events on the top-end of the upper wind, since the inclination of
the model is not exactly 90$^\circ$.

\section{Discussion}
Polarization mapping is a unique tool to visualize the morphological differences between different reprocessing media. In the context of the model for 
quasars presented by \citet{Elvis2000}, it appears that the polarization signatures show a wide panel of geometries, revealing the outermost part of the structures. 
In comparison with a usual, dusty torus \citep{Marin2012}, type-1 views are rather similar between a torus and a disk-born wind. The funnel of the model is 
highly irradiated, while the extended parts of the regions are less impacted by the input radiation. Looking at an intermediate viewing angle, a dusty wind 
becomes more similar to a double-cone geometry in terms of flux repartition. The small width of the outflow allows radiation to escape from the inner regions 
and the polarized flux is stronger than in type-1 inclinations. The main difference between a torus and a disk-born wind is seen at equatorial, type-2 viewing angles. 
The central part of the map shows the brightest flux, associated with intermediate degrees of polarization. The rest of the model is nearly invisible due to absorption.

If we think that a dusty outflow fits in with the observed presence of dusty NLR clouds at larger distances from the irradiation source, then it is hard 
to reconcile the resulting polarization maps with the observations of IC~5063 \citep{Morganti2007}. This southern Seyfert-2 galaxy shows an ionized structure 
extending out to 15~kpc in a remarkable, X-shaped morphology \citep{Morganti2003}. In Marin \& Goosmann (2013, in press), we found that a bi-phased outflow
well reproduce both the polarization degrees and the geometrical X shape. Here, we complementary demonstrate that a wind uniquely filled with dust grains
cannot reproduce the flux morphology of IC~5063 as the hollow outflows mostly absorb the inner radiation.

\section{Summary and conclusions}
As a complementary work to the spectrophotometric modeling of a dusty outflow (Marin \& Goosmann, submitted), based on morphological constraints presented
in \citet{Elvis2000}, we modeled the optical/UV polarization maps of a theoretical, dusty model. We found that the resulting maps present various 
geometries, depending on the observer's viewing angle. While for a polar inclination, a dusty disk-born outflow can be mistaken for a regular, obscuring torus,
intermediate and edge-on views show rather different results. New observational campaigns, using polarimetric imagers would greatly help to disentangle
between a hydrostatic torus and a dynamic outflow, but our present conclusions show that a pure dusty outflow hardly reproduces the observational data.
As suggested in Marin \& Goosmann (2013, in press), refining the bending and divergence angles of the flow could lead to better simulations. The academic model
presented here should not be taken as an actual model to replace a toroidal region, but as a step toward an upgrade of the model by \citet{Elvis2000}.

Future work, focusing on broadband polarimetric signature, velocity fields and absorption/emission features will be conducted to push forward the conclusions 
drawn in Marin \& Goosmann (2013, in press).

\bibliographystyle{aa} 
\bibliography{marin} 

\end{document}